\documentclass[twocolumn,english]{revtex4}
\usepackage[T1]{fontenc}
\usepackage[latin9]{inputenc}
\usepackage{color}
\usepackage{array}
\usepackage{float}
\usepackage{textcomp}
\usepackage{multirow}
\usepackage{amsmath}
\usepackage{amssymb}

\makeatletter

\providecommand{\tabularnewline}{\\}

\@ifundefined{textcolor}{}
{%
 \definecolor{BLACK}{gray}{0}
 \definecolor{WHITE}{gray}{1}
 \definecolor{RED}{rgb}{1,0,0}
 \definecolor{GREEN}{rgb}{0,1,0}
 \definecolor{BLUE}{rgb}{0,0,1}
 \definecolor{CYAN}{cmyk}{1,0,0,0}
 \definecolor{MAGENTA}{cmyk}{0,1,0,0}
 \definecolor{YELLOW}{cmyk}{0,0,1,0}
}

\makeatother

\usepackage{babel}
\begin{document}
\title{Analytic gradients for spin multiplets in natural orbital functional
theory\bigskip{}
}
\author{Ion Mitxelena$^{1}$}
\author{Mario Piris$^{1,2}$}
\email{mario.piris@ehu.eus}

\address{$^{1}$Donostia International Physics Center (DIPC), 20018 Donostia,
Euskadi, Spain. \\
$^{2}$Kimika Fakultatea, Euskal Herriko Unibertsitatea (UPV/EHU)
and Basque Foundation for Science (IKERBASQUE), 48013 Bilbao, Euskadi,
Spain.\bigskip{}
}
\begin{abstract}
Analytic energy gradients with respect to nuclear motion are derived
for non-singlet compounds in the natural orbital functional theory.
We exploit the formulation for multiplets in order to obtain a simple
formula valid for any many-electron system in its ground mixed state
with a total spin $S$ and all possible spin projection $S_{z}$ values.
We demonstrate that the analytic gradients can be obtained without
resorting to linear response theory or involving iterative procedures.
It is required a single evaluation, so integral derivatives can be
computed on-the-fly along the calculation and thus improve the effectiveness
of screening by the Schwarz inequality. Results for small and medium
size molecules with many spin multiplicities are shown. Our results are 
compared with experimental data and accurate theoretical equilibrium geometries.\bigskip{}

\bigskip{}
\end{abstract}
\maketitle

\section{Introduction}

Non-singlet systems are a non-trivial problem for reduced density
matrix (RDM) based electronic structure methods. Most commonly used
procedures focus on the high-spin component \cite{Leiva2007,Piris2009,Rohr2011,Quintero2019}
or break spin symmetry \cite{alcoba-2019}. In either of these cases,
it is not possible to describe the degeneracy of the energy levels
with respect to the spin projection ($S_{z}$). In fact, a single
energy value is available if we focus only on the high-spin component,
whereas the total spin $S$ is not a proper quantum number whenever
the unrestricted formalism is used. The correct spin symmetry can
be recovered by \textit{a posteriori} projection \cite{scuseria-2017},
however, the latter applies only to the energy. Alternatively, Alcoba
and co-workers have developed \cite{alcoba-2008} a purification procedure
that simultaneously retrieves the N- and $S$-representability
of the two-particle RDM (2RDM). Unfortunately, the latter implies
an added complexity since different spin symmetries require explicit
conditions to be imposed. An incorrect treatment of the spin symmetry
may have dramatic consequences on the molecular equilibrium geometries.
Consequently, it is of fundamental importance to compute energy gradients
preserving the spin symmetry of the system.

Recently \cite{Piris2019}, a new formulation of the first-order RDM
(1RDM) functional theory in its spectral representation, the so-called
natural orbital functional (NOF) theory (NOFT), was proposed for multiplet
states. Then the 1RDM is expressed by the natural orbitals (NOs) and
their occupation numbers (ONs). This approach relies on considering
the reconstruction of the ensemble 2RDM in terms of the ONs to describe
a many-electron system in its ground mixed state with a total spin
$S$ and all possible $S_{z}$ values. It turns out that the expected
value of $\hat{S}_{z}$ for the whole ensemble is zero, therefore
the spin-restricted theory can be adopted although the total spin
is non-zero. A particularly relevant consequence of considering the
spin multiplet is that the energy derivatives of non-singlet compounds
can be computed in complete analogy to their calculation in singlet
states.

The efficient computation of analytic energy gradients is now under
intense development in RDM based methods \cite{maradzike,mullinax:2019:6164,VALENTINE2017300,mazziotti-gradients,Sokolovdcft,MITXELENA2019}.
In NOFT, the general formulation of analytic gradients was presented
in Ref. \cite{Mitxelena2017}, where equilibrium geometries of molecules
in singlet states were efficiently calculated. In fact, it was demonstrated
that the analytic gradients can be obtained by simple evaluation without
resorting to linear response theory or involving iterative procedures.
In this article, we demonstrate that the latter also applies to non-singlet
compounds in NOFT as long as the entire multiplet is considered. Consequently,
equilibrium geometries of molecular systems with any spin value can
be easily obtained preserving the total spin.

This article is organized as follows. We begin in section \ref{sec:Theory}
with a brief review of the analytic gradients with respect to nuclear
displacements in the context of NOFT. Then, we expose the details
of our approach for mixed quantum states with nonzero total spin.
Computational aspects related to an efficient gradient computation
are also discussed. Results for molecular systems involving many spin
multiplicities are shown in section \ref{sec:Equilibrium-geometries}.
The latter includes a comparison with respect to experimental data,
as well as standard electronic structure methods such as Hartree-Fock
(HF) or coupled-cluster including singles and doubles excitations
with perturbative triples {[}CCSD(T){]}. At the end we highlight the
main conclusions of this paper.

\section{\label{sec:Theory}Theory}

\subsection{Analytic gradients in NOFT\label{subsec: Analytic-gradients-in-NOFT}}

In the Born-Oppenheimer approximation, the total energy of an N-electron
molecule can be cast as the sum of the nuclear ($E_{nuc}$) and electronic
($E_{el}$) energies,
\begin{equation}
E=E_{nuc}+E_{el}={\displaystyle \sum_{A<B}}\frac{Z_{A}Z_{B}}{R_{AB}}+E_{el}.\label{Etotal}
\end{equation}

$Z_{A}$ represents the atomic number of nucleus $A$, and $R_{AB}$
is the distance between nuclei $A$ and $B$. $E_{el}$\textcolor{black}{{}
}is an exactly and explicitly known functional of the 1RDM ($\Gamma$)
and 2RDM ($D$), 
\begin{equation}
E_{el}=\sum\limits _{ik}\Gamma_{ki}\mathcal{H}_{ki}+\sum\limits _{ijkl}D_{klij}\left\langle kl|ij\right\rangle ,\label{Eelec_0}
\end{equation}
resulted from a non-relativistic Hamiltonian without spin coordinates,
namely,
\begin{equation}
\hat{H}=\sum\limits _{ik}\mathcal{H}_{ki}\hat{a}_{k}^{\dagger}\hat{a}_{i}+\frac{1}{2}\sum\limits _{ijkl}\left\langle kl|ij\right\rangle \hat{a}_{k}^{\dagger}\hat{a}_{l}^{\dagger}\hat{a}_{j}\hat{a}_{i}.\label{Ham}
\end{equation}

$\mathcal{H}_{ki}$ are matrix elements of the one-electron part of
the Hamiltonian involving the kinetic and potential energy operators,
whereas $\left\langle kl|ij\right\rangle $ are the two-electron integrals
of the Coulomb interaction. $\hat{a}_{i}^{\dagger}$ and $\hat{a}_{i}$
are the familiar fermion creation and annihilation operators associated
with the complete orthonormal spin-orbital set $\left\{ \left|i\right\rangle \right\} $.
In this context, the ground state is a multiplet, i.e., an ensemble
of states that allows all possible spin projections. For a given $S$,
there are $\left(2S+1\right)$ energy degenerate eigenvectors $\left|SM_{s}\right\rangle $,
so the ground state is defined by the N-particle density matrix statistical
operator of all equiprobable pure states:
\begin{equation}
\mathfrak{\hat{D}}={\displaystyle \dfrac{1}{2S+1}{\displaystyle {\textstyle {\displaystyle \sum_{M_{s}=-S}^{S}}}}\left|SM_{s}\right\rangle \left\langle SM_{s}\right|.}\label{DM}
\end{equation}
From Eq. (\ref{Ham}), it follows that matrix elements of the RDMs
are
\begin{equation}
\begin{array}{c}
\Gamma_{ki}={\displaystyle \dfrac{1}{2S+1}{\textstyle {\displaystyle \sum_{M_{s}=-S}^{S}}}}\left\langle SM_{s}\right|\hat{a}_{k}^{\dagger}\hat{a}_{i}\left|SM_{s}\right\rangle ,\\
D_{klij}={\displaystyle {\textstyle {\displaystyle \dfrac{1}{2\left(2S+1\right)}\sum_{M_{s}=-S}^{S}}}}\left\langle SM_{s}\right|\hat{a}_{k}^{\dagger}\hat{a}_{l}^{\dagger}\hat{a}_{j}\hat{a}_{i}\left|SM_{s}\right\rangle .
\end{array}
\end{equation}
The Löwdin's normalization is used, therefore, the traces of the 1RDM
and 2RDM are equal to the number of electrons and the number of electron
pairs, respectively. 

To construct an approximate NOF, we employ the representation where
the 1RDM is diagonal ($\Gamma_{ki}=n_{i}\delta_{ki}$). Restriction
on the ONs to the range $0\leq n_{i}\leq1$ represents a necessary
and sufficient condition for ensemble N-representability of the 1RDM
\cite{Coleman1963}. We further assume that all NOs $\left\{ \phi_{i}\left(\mathbf{x}\right)=\left\langle \mathbf{x}|i\right\rangle \right\} $
are real and expand them in a fixed basis set,
\begin{equation}
\phi_{i}\left(\mathbf{x}\right)=\sum_{\upsilon}\mathcal{C}_{\upsilon i}\zeta_{\upsilon}\left(\mathbf{x}\right).\label{lcao}
\end{equation}
In Eq. (\ref{lcao}), ${\bf x\equiv}\left({\bf r,s}\right)$ stands
for the combined spatial and spin coordinates, ${\bf r}$ and ${\bf s}$,
respectively. Accordingly, the electronic energy (\ref{Eelec_0})
can be rewritten as
\begin{equation}
E_{el}={\displaystyle \sum_{\mu\upsilon}\Gamma_{\mu\upsilon}\mathcal{H}_{\mu\upsilon}+\sum_{\mu\upsilon\eta\delta}D_{\mu\eta\upsilon\delta}\left\langle \mu\eta|\upsilon\delta\right\rangle },\label{Eel_2}
\end{equation}
where $\Gamma_{\mu\upsilon}$ and $D_{\mu\eta\upsilon\delta}$ are
the RDMs in the atomic orbital (AO) representation,
\begin{equation}
\begin{array}{c}
\Gamma_{\mu\upsilon}={\displaystyle \sum_{i}}n_{i}\mathcal{C}_{\mu i}\mathcal{C}_{\upsilon i},\\
D_{\mu\eta\upsilon\delta}=\sum\limits _{klij}D_{klij}\mathcal{C}_{\mu k}\mathcal{C}_{\eta l}\mathcal{C}_{\upsilon i}\mathcal{C}_{\delta j}.
\end{array}\label{RDM_ao_1}
\end{equation}
Then, the derivative of the total energy (\ref{Etotal}) with respect
to the coordinate $x$ of nucleus $A$ is given by \cite{Mitxelena2017}
\begin{equation}
\begin{array}{c}
{\displaystyle \frac{dE}{dx_{A}}}={\displaystyle \frac{\partial E_{nuc}}{\partial x_{A}}}+{\displaystyle \sum_{\mu\upsilon}}\Gamma_{\mu\upsilon}\dfrac{\partial\mathcal{H}_{\mu\upsilon}}{\partial x_{A}}+\\
{\displaystyle \sum_{\mu\eta\upsilon\delta}}D_{\mu\eta\upsilon\delta}\dfrac{\partial\left\langle \mu\eta|\upsilon\delta\right\rangle }{\partial x_{A}}-{\displaystyle \sum_{\mu\upsilon}}\lambda_{\mu\upsilon}\dfrac{\partial\mathcal{S_{\mu\upsilon}}}{\partial x_{A}}.
\end{array}\label{NOF-analy-grads}
\end{equation}
The first term of Eq. (\ref{NOF-analy-grads}) is the derivative of
the nuclear energy, the second represents the negative Hellmann-Feynman
force, while the third contains the explicit derivatives of two-electron
integrals. The last term in Eq. (\ref{NOF-analy-grads}), known as
the density force\textit{ }\cite{pulay}\textit{,} arises from the
implicit dependence of the orbital coefficients $\left\{ \mathcal{C}_{\upsilon i}\right\} $
on geometry. Recall that the implicit dependence of ONs on geometry
does not contribute to analytic gradients since $E_{el}$ is stationary with respect to variations 
in all of the ONs (see Eq. (17) in Ref. \cite{Mitxelena2017}).

In Eq. (\ref{NOF-analy-grads}), $S_{\mu\upsilon}=\left\langle \mu|\upsilon\right\rangle $
is the overlap matrix in the AO representation, whereas the Lagrange
multipliers $\lambda_{\mu\upsilon}$ are calculated as
\begin{equation}
\begin{array}{c}
{\displaystyle {\displaystyle \lambda_{\mu\upsilon}=\sum_{ij}\mathcal{C}_{\mu j}}\lambda_{ji}\mathcal{C}_{\upsilon i},}\\
\lambda_{ji}=n_{i}\mathcal{H}_{ji}+2\,{\displaystyle \sum_{klm}}D_{klim}\left\langle kl|jm\right\rangle .
\end{array}\label{lagmul}
\end{equation}

It is worth noting that all derivatives in Eq. (\ref{NOF-analy-grads})
have an explicit dependence on the nuclear coordinate $x_{A}$. Consequently,
the analytic gradients $dE/dx_{A}$ can be obtained by a single evaluation
since NOs and ONs corresponding to perturbed geometries are not required.
Following Ref. \cite{Mitxelena2017}, two-electron integral derivatives
$\partial\left\langle \mu\eta|\upsilon\delta\right\rangle /\partial x_{A}$
remain the bottleneck of gradient evaluation, so they are computed
on-the-fly in order to efficiently apply the Schwarz' screening \cite{horn}
with corresponding savings of storage and computing times. Additionally,
the present implementation in the DoNOF computer program \cite{Piris2020}
allows to carry out the geometry optimization procedure ruled by Eq.
(\ref{NOF-analy-grads}) using the conjugate gradient or L-BFGS algorithms.

\subsection{\label{sec:Natural-Orbital-Functional}Natural Orbital Functional
for Multiplets}

To construct an approximate NOF, the second term of the electronic
energy, which is an explicit functional of the 2RDM, must be reconstructed
from the 1RDM. In our case, we reconstruct the ensemble 2RDM from
the ONs. We neglect any explicit dependence of $D$ on the NOs themselves
given that the energy functional already has a strong dependence on
the latter via the one- and two-electron integrals. In the following,
the reconstruction $D[n_{i},n_{j},n_{k},n_{l}]$ leading to spin-multiplets
is briefly described. For further details, the reader should consult
Ref. \cite{Piris2019}.

In this work, we shall use the static version of our latest functional,
namely PNOF7s. This functional approximation was introduced \cite{Piris2018b}
to provide the reference NOF in the second-order Möller-Plesset perturbation
method (NOF-MP2) that allows the inter-pair dynamic correlation to
be recovered. It is worth noting that PNOF7s satisfies electron pairing
restrictions \cite{Piris2018a}, but only takes into account the inter-pair
static electron correlation, thus preventing the ONs and NOs from
experiencing spurious inter-pair correlation in the dynamic correlation
domains characteristic of molecular equilibrium regions.

Consider $\mathrm{N_{I}}$ single electrons determine the spin $S$
of the system, and the rest of electrons ($\mathrm{N_{II}}=\mathrm{N-N_{I}}$)
are spin-paired, so that all spins corresponding to $\mathrm{N_{II}}$
electrons provide a zero spin. Next, focus on the mixed state of highest
multiplicity: $2S+1=\mathrm{N_{I}}+1,\,S=\mathrm{N_{I}}/2$. For the
ensemble of pure states $\left\{ \left|SM_{s}\right\rangle \right\} $,
we have that
\begin{equation}
\text{\textlangle}\hat{S}_{z}\text{\textrangle}=\frac{1}{\mathrm{N_{I}}+1}{\textstyle {\displaystyle \sum_{M_{s}=-\mathrm{N_{I}}/2}^{\mathrm{N_{I}}/2}}M_{s}}=0.\label{SZ0}
\end{equation}

Eq. (\ref{SZ0}) implies that the expected value of $\hat{S}_{z}$
for the whole ensemble is zero. This result is of transcendental importance
for the present work. In fact, the spin-restricted theory can be adopted
although the total spin of the system is non-zero. Consequently, we
can use a single set of orbitals for $\alpha$ and $\beta$ spins.
All the spatial orbitals will be then doubly occupied in the ensemble,
so that occupancies for particles with $\alpha$ and $\beta$ spins
are equal: $n_{p}^{\alpha}=n_{p}^{\beta}=n_{p}.$ On the other hand,
the RDMs (\ref{RDM_ao_1}) and Lagrange multipliers (\ref{lagmul})
become
\begin{equation}
\begin{array}{c}
\Gamma_{\mu\upsilon}=2{\displaystyle \sum_{p}}n_{p}\mathcal{C}_{\mu p}\mathcal{C}_{\upsilon p},\\
D_{\mu\eta\upsilon\delta}=2\sum\limits _{pqrt}(D_{pqrt}^{\alpha\alpha}+D_{pqrt}^{\alpha\beta})\mathcal{C}_{\mu p}\mathcal{C}_{\eta q}\mathcal{C}_{\upsilon r}\mathcal{C}_{\delta t},\\
{\displaystyle \lambda_{\mu\upsilon}=2\sum_{pq}}\mathcal{C}_{\mu q}\lambda_{qp}\mathcal{C}_{\upsilon p},\\
\lambda_{qp}=n_{p}\mathcal{H}_{qp}+4{\displaystyle \sum_{rtu}}(D_{rtpu}^{\alpha\alpha}+D_{rtpu}^{\alpha\beta})\left\langle rt|qu\right\rangle .
\end{array}\label{RDM_ao_2}
\end{equation}

To achieve the gradients (\ref{NOF-analy-grads}), we obviously need
specific approximations for the 2RDM spin-components. Let us divide
the orbital space into two subspaces: $\Omega=\Omega_{\mathrm{I}}\oplus\Omega_{\mathrm{II}}$.
$\Omega_{\mathrm{II}}$ is composed of $\mathrm{N_{II}}/2$ mutually
disjoint subspaces $\Omega{}_{g}$. Each of which contains one orbital
$\left|g\right\rangle $ with $g\leq\mathrm{N_{II}}/2$, and $\mathrm{N}_{g}$
orbitals $\left|p\right\rangle $ with $p>\mathrm{N_{II}}/2$, namely,
\begin{equation}
\Omega{}_{g}=\left\{ \left|g\right\rangle ,\left|p_{1}\right\rangle ,\left|p_{2}\right\rangle ,...,\left|p_{\mathrm{N}_{g}}\right\rangle \right\} .\label{OmegaG}
\end{equation}
Taking into account the spin, the total occupancy for a given subspace
$\Omega{}_{g}$ is 2, which is reflected in the following sum rule:
\begin{equation}
\sum_{p\in\Omega_{\mathrm{II}}}n_{p}=n_{g}+\sum_{i=1}^{\mathrm{N}_{g}}n_{p_{i}}=1,\quad g=1,2,...,\frac{\mathrm{N_{II}}}{2}.\label{sum1}
\end{equation}
In general, $\mathrm{N}_{g}$ may be different for each subspace,
but it should be sufficient for the description of each electron pair.
In this work, $\mathrm{N}_{g}$ is equal to a fixed number for all
subspaces $\Omega{}_{g}\in\Omega_{\mathrm{II}}$. The maximum possible
value of $\mathrm{N}_{g}$ is determined by the basis set used in
calculations. From (\ref{sum1}), it follows that
\begin{equation}
2\sum_{p\in\Omega_{\mathrm{II}}}n_{p}=2\sum_{g=1}^{\mathrm{N_{II}}/2}\left(n_{g}+\sum_{i=1}^{\mathrm{N}_{g}}n_{p_{i}}\right)=\mathrm{N_{II}}.\label{sumNpII}
\end{equation}

Here, the notation $p\in\Omega_{\mathrm{II}}$ represents all the
indexes of $\left|p\right\rangle $ orbitals belonging to $\Omega_{\mathrm{II}}$.
It is important to recall that orbitals belonging to each subspace
$\Omega_{g}$ vary along the optimization process until the most favorable
orbital interactions are found. Therefore, orbitals do not remain
fixed in the optimization process, they adapt to the problem.

Similarly, $\Omega_{\mathrm{I}}$ is composed of $\mathrm{N_{I}}$
mutually disjoint subspaces $\Omega{}_{g}$. In contrast to $\Omega_{\mathrm{II}}$,
each subspace $\Omega{}_{g}\in\Omega_{\mathrm{I}}$ contains only
one orbital $g$ with $2n_{g}=1$. It is worth noting that each orbital
is completely occupied individually, but we do not know whether the
electron has $\alpha$ or $\beta$ spin: $n_{g}^{\alpha}=n_{g}^{\beta}=n_{g}=1/2$.
It follows that
\begin{equation}
2\sum_{p\in\Omega_{\mathrm{I}}}n_{p}=2\sum_{g=\mathrm{N_{II}}/2+1}^{\mathrm{N_{II}}/2+\mathrm{N_{I}}}n_{g}=\mathrm{N_{I}}.\label{sumNpI}
\end{equation}
Taking into account Eqs. (\ref{sumNpII}) and (\ref{sumNpI}), the
trace of the 1RDM is verified equal to the number of electrons:
\begin{equation}
2\sum_{p\in\Omega}n_{p}=2\sum_{p\in\Omega_{\mathrm{II}}}n_{p}+2\sum_{p\in\Omega_{\mathrm{I}}}n_{p}=\mathrm{N_{II}}+\mathrm{N_{I}}=\mathrm{\mathrm{N}}.\label{norm}
\end{equation}

To guarantee the existence of an N-electron system compatible with
a NOF, we must observe the N-representability conditions \cite{Mazziotti2012a}
on the reconstructed 2RDM \cite{Piris2018}. In electron-pairing-based
NOFs, the employment of these constraints leads to divide the matrix
elements of $D$ into intra- and inter-subspace contributions. The
intra-subspace blocks only involves intrapair $\alpha\beta$-contributions
of orbitals belonging to $\Omega_{\mathrm{II}}$,
\begin{equation}
\begin{array}{c}
D_{pqrt}^{\alpha\beta}={\displaystyle \frac{\Pi_{pr}}{2}}\delta_{pq}\delta_{rt}\delta_{p\Omega_{g}}\delta_{r\Omega_{g}}\;(g\leq\frac{N_{\mathrm{II}}}{2}),\quad\\
\\
\Pi_{pr}=\left\{ \begin{array}{c}
\sqrt{n_{p}n_{r}}\qquad p=r\textrm{ or }p,r>\frac{N_{\mathrm{II}}}{2}\\
-\sqrt{n_{p}n_{r}}\qquad p=g\textrm{ or }r=g\qquad\;
\end{array},\right.\\
\\
\delta_{p\Omega_{g}}=\left\{ \begin{array}{c}
1\,,\:p\in\Omega_{g}\\
0\,,\:p\notin\Omega_{g}
\end{array},\right.
\end{array}\label{intra}
\end{equation}
whereas inter-subspace contributions ($\Omega_{f}\neq\Omega{}_{g}$)
for the spin-parallel matrix elements are HF like,
\begin{equation}
D_{pqrt}^{\sigma\sigma}={\displaystyle \frac{n_{p}n_{q}}{2}}\left(\delta_{pr}\delta_{qt}-\delta_{pt}\delta_{qr}\right)\delta_{p\Omega_{f}}\delta_{q\Omega_{g}},\:{}_{\sigma=\alpha,\beta},\label{interaa}
\end{equation}
and the spin-anti-parallel blocks take the form
\begin{equation}
\begin{array}{c}
D_{pqrt}^{\alpha\beta}={\displaystyle {\displaystyle \frac{n_{p}n_{q}}{2}}}\delta_{pr}\delta_{qt}\delta_{p\Omega_{f}}\delta_{q\Omega_{g}}-{\displaystyle \frac{\Phi_{p}\Phi_{r}}{2}}\delta_{p\Omega_{f}}\delta_{r\Omega_{g}}\cdot\\
\\
\cdot\left\{ \begin{array}{c}
\delta_{pq}\delta_{rt}\quad f\leq\frac{\mathrm{N_{II}}}{2}\textrm{ or }g\leq\frac{\mathrm{N_{II}}}{2}\\
\quad\delta_{pt}\delta_{qr}\qquad\frac{\mathrm{N_{II}}}{2}<f,g\leq\mathrm{N}_{\Omega}\quad
\end{array}.\right.
\end{array}\label{interab}
\end{equation}

Eqs. (\ref{intra})-(\ref{interab}), together with $\Phi_{p}=2n_{p}(1-n_{p})$
in Eq. (\ref{interab}), result in PNOF7s for multiplets. It is not
difficult to verify \cite{Piris2019} that the reconstruction (\ref{intra})-(\ref{interab})
leads to $\mathrm{<}\hat{S}^{2}\mathrm{>}=S\left(S+1\right)$ with
$S=\mathrm{N_{I}}/2$. Therefore, the PNOF7s equilibrium geometries
obtained in section \ref{sec:Equilibrium-geometries} should be exempt from
spin contamination effects.

Replacing the 2RDM components described by (\ref{intra})-(\ref{interab})
in Eq. (\ref{RDM_ao_2}), we have all the necessary elements to obtain
the PNOF7s analytic gradients through the Eq. (\ref{NOF-analy-grads}).
Then, we are in position to calculate PNOF7s equilibrium geometries
for multiplets by unconstrained optimizations with respect to nuclear
coordinates.

\section{\label{sec:Equilibrium-geometries}Equilibrium geometries}

In this section, we study the equilibrium geometries of 32 molecules
in their multiplet ground states. We have included exclusively doublets
and triplets spin states. All results corresponding to NOF calculations
were carried out using DoNOF \cite{Piris2020}, our open-source implementation
of NOF-based methods for quantum chemistry. The correctness of the analytic 
gradients was verified numerically. The correlation-consistent
polarized valence triple-zeta (cc-pVTZ) basis set developed by Dunning
and co-workers \cite{Dunning1989} was shown \cite{Mitxelena2017}
to be suitable for PNOF geometry optimizations in singlet states.
Consequently, we employ the cc-pVTZ basis set in all multiplet calculations.

For comparison, we have included high-quality empirical equilibrium structures 
obtained from least-squares fits involving experimental rotational constants and 
theoretical vibrational corrections. It is worth noting that we do not intend to 
reproduce the experimental data in this paper, as this requires going to the complete 
basis set limit. In order to eliminate the effect of basis set incompleteness error 
on the results, we compare our equilibrium geometries with the geometries obtained 
by other theoretical methods with the same basis set cc-pVTZ, namely HF, MP2, and CCSD(T). 
Geometries calculated with these theoretical methods as well as experimental data 
were obtained from the NIST Computational Chemistry Comparison and Benchmark Database \cite{nist}.

Hereafter, we set $\mathrm{N}_{g}=3$ so each subspace $\Omega{}_{g}\in\Omega_{\mathrm{II}}$
is composed of one strongly-occupied orbital coupled with three weakly-occupied
orbitals. In fact, $\mathrm{N}_{g}=3$ should be sufficient to describe
the intrapair correlation of molecules studied here. It has already
been shown \cite{Mitxelena2017} that PNOF equilibrium geometries
do not depend significantly on $\mathrm{N}_{g}$. As starting geometries
in our optimizations, we take those previously obtained by the HF
method. Harmonic vibrational frequency analyzes were performed for
all systems using numerical differentiation of analytic gradients
\cite{MITXELENA2019}. In all cases it was verified that a minimum
of the geometry had been reached.

\begin{table}[H]
\caption{\label{estad-diatomics}Mean signed (MSE), mean unsigned (MUE), maximum (MAX) 
and root-mean-squared (RMSE) errors for the bond distances ($\textrm{Å}$) of diatomic
molecules regarding experimental data. See appendix for the equilibrium distance
data.}

\centering{}\bigskip{}
\begin{tabular}{c|cccc}
 & PNOF7s & HF & MP2 & CCSD(T)\tabularnewline
\hline 
 MSE  & $\,$ 0.000 & -0.009     & -0.001     & $\,$ 0.007 \tabularnewline
 MUE  & $\,$ 0.011 & $\,$ 0.020 & $\,$ 0.016 & $\,$ 0.013 \tabularnewline
 MAX  & -0.039     & -0.049     & -0.071     & -0.066     \tabularnewline
 RMSE & $\,$ 0.015 & $\,$ 0.024 & $\,$ 0.023 & $\,$ 0.019 \tabularnewline
\end{tabular}
\end{table}

Table \ref{estad-diatomics} shows the mean signed (MSE), mean unsigned
(MUE), maximum (MAX) and root-mean-squared (RMSE) errors corresponding to 
the equilibrium bond distances obtained for 24 diatomic molecules at the PNOF7s, 
HF, MP2, and CCSD(T) levels of theory. Bond distance errors were calculated as 
the differences between theoretical results and experimental lengths. The data used
to calculate these errors can be found in Table 1 of the Appendix.
From Table \ref{estad-diatomics}, we observe that PNOF7s produces the
most accurate bond distances for molecules consisting of two atoms,
closely followed by CCSD(T) and MP2 methods.

Table \ref{estad-poly} displays the MSE, MUE, MAX, and RMSE in bond distances corresponding 
to molecules composed of more than two atoms. The theoretical and experimental
equilibrium bond distances for the 8 polyatomic molecules studied
can be found in Table 2 of the Appendix. In contrast with the results
obtained for diatomic molecules, the lowest MUE and RMSE are obtained with CCSD(T), while
HF continues giving the largest errors. According to the values obtained
for the MSE, HF significantly underestimates the bond
distances, whereas CCSD(T) tends to slightly overestimate them. In
the case of PNOF7s and MP2, the very small mean signed error values
reveal that the bond distances are well compensated, making cc-pVTZ
adequate for calculating equilibrium geometries using these methods.

\begin{table}[H]
\caption{\label{estad-poly}Mean signed (MSE), mean unsigned (MUE), maximum (MAX) 
and root-mean-squared (RMSE) errors for bond distances ($\textrm{Å}$) of polyatomic 
molecules regarding experimental data. See appendix for the equilibrium distance
data.}

\centering{}\bigskip{}
\begin{tabular}{c|cccc}
 & PNOF7s & HF & MP2 & CCSD(T)\tabularnewline
\hline 
 MSE  & $\,$ 0.000 & -0.013     & -0.001     & 0.007 \tabularnewline
 MUE  & $\,$ 0.018 & $\,$ 0.026 & $\,$ 0.018 & 0.014 \tabularnewline
 MAX  & -0.054     & -0.054     & $\,$ 0.062 & 0.069 \tabularnewline
 RMSE & $\,$ 0.025 & $\,$ 0.030 & $\,$ 0.023 & 0.023 \tabularnewline
\end{tabular}
\end{table}

Compared to the MUE and RMSE obtained for the bond distances of diatomic molecules, 
it is worth noting that the PNOF7s errors get larger as the system increases in 
polyatomic molecules. Recall that in NOFT based on electron pairs \cite{Piris2018a}, 
the intrapair electron correlation is appropriately captured, but a significant
part of the correlation between electron pairs is lost. In the case
of PNOF7s, the static interpair correlation is well described, but
not the dynamic one. The lack of interpair dynamic correlation becomes
significant when the number of electron pairs increases, so a worsening
in the description of the geometries is expected.

The dynamic correlation can be incorporated using perturbation corrections,
which led to the NOF-MP2 method \cite{Piris2017,Piris2018b}. Recently,
it has been shown that a more accurate description of the potential
energy surface (PES) can be obtained (eg, See Fig. 4 in Ref. \cite{Piris2020})
using NOF-MP2. Unfortunately, obtaining the energy gradients corresponding
to methods based on the MP2 formulation implies additional complexity
\cite{mp2-grads-2019}, since it is necessary to iteratively solve
coupled perturbed equations \cite{Azhary}. Therefore, the analytic
gradients of the NOF-MP2 method are beyond the scope of this work.

Overall, the PNOF7s MSE, MUE and RMSE obtained for bond distances are 0.000 
$\textrm{Å}$, 0.014 $\textrm{Å}$ and 0.020 $\textrm{Å}$, respectively. Therefore, the
conclusions previously obtained \cite{MITXELENA2019} for spin compensated
systems hold in the present study. On the one hand, bond distances
obtained at the PNOF7s/cc-pVTZ level of theory are in good agreement
with the experimental data and, on the other, PNOF7s produces slightly
larger errors than the CCSD(T) method. In fact, the errors
corresponding to CCSD(T) read as MUE = 0.013 $\textrm{Å}$ and RMSE = 0.020 $\textrm{Å}$. 
However, we must recall that CCSD(T) requires solving
coupled perturbed equations to compute the energy gradients and introduces
spin contamination effects due to the use of the spin unrestricted
approach in the case of non-zero spin systems.

\begin{table}[H]
\caption{\label{estad-angles}Mean signed (MSE), mean unsigned (MUE), maximum (MAX) 
and root-mean-squared (RMSE) errors for bond angles (degrees) of polyatomic molecules 
regarding experimental data. See appendix for the equilibrium angle data.}

\centering{}\bigskip{}
\begin{tabular}{c|cccc}
 & PNOF7s & HF & MP2 & CCSD(T)\tabularnewline
\hline 
 MSE  & -0.2     & 1.3 & 0.2 & 0.4 \tabularnewline
 MUE  & $\,$1.4  & 2.2 & 1.2 & 1.2 \tabularnewline
 MAX  & -3.1     & 7.2 & 4.5 & 4.9  \tabularnewline
 RMSE & $\,$ 1.8 & 3.1 & 1.9 & 1.9 \tabularnewline
\end{tabular}
\end{table}

Table \ref{estad-angles} shows the MSE, MUE, MAX, and RMSE in bond angles 
corresponding to the 8 studied polyatomic molecules. The data used to calculate 
these errors can be found in Table 3 of the Appendix. Despite imposing no symmetry
on PNOF7s calculations, it is worth noting that DoNOF mantains the
point group symmetry of the starting HF molecular equilibrium geometries.
According to the values reported in Table \ref{estad-angles}, bond
angles obtained using PNOF7s are in good agreement with those obtained
at the MP2 and CCSD(T) levels of theory, and thereby imply a significant
improvement with respect to HF results. In contrast to the results
obtained for bond lengths, PNOF7s produces a non-negligible MSE of -0.2 degrees, 
with a slight tendency to underestimate bond angles. Nevertheless, together with the MSE 
obtained for the bond distances, Table \ref{estad-angles} shows that the cc-pVTZ
basis set conforms to PNOF7s in order to obtain accurate equilibrium
geometries.

\section{Closing Remarks}

In this work, it has been demonstrated that the analytic energy gradients
of non-singlet compounds in their ground state can be obtained analogously
to singlets by using the NOF approach for multiplets. The latter describes
a mixed quantum state with all possible spin projections, so the energy
degeneration with respect to these projections is adequately accounted
for. The expectation value of $\hat{S}^{2}$ is correct, so the obtained 
equilibrium geometries should be exempt from spin contamination effects. The previously
developed formulas to compute gradients for singlets are valid for
non-zero values of total spin. Thus, the NOF gradients are easily
computed by a single evaluation and without solving coupled perturbed
equations.

Geometry optimization was performed for 32 molecules with spin multiplicities
equal to 2 and 3. In all cases, it was confirmed that all the harmonic
vibrational frequencies were positive, so it was corroborated that
we had reached a minimum in the potential energy surfaces. In most
cases, PNOF7s geometries are comparable to those obtained at MP2 and
CCSD(T) levels of theory, so a significant improvement with respect
to HF results is obtained. Since the mean signed error is nearly zero, and
the mean unsigned and root-mean-squared errors are close to those obtained with 
the accurate CCSD(T) method, the cc-pVTZ basis set appears to be suitable for 
calculating equilibrium geometries with PNOF7s.

In view of the results shown throughout this work, PNOF7s is an efficient
method to calculate the equilibrium geometries of non-singlet systems. 
Furthermore, approximate NOFs deal with static correlation from the outset and do not require
prior knowledge of the system or parameterization techniques. This
makes PNOF7s a method of choice to be used in problems related to
energy gradients and systems where an accurate description of the
non-dynamic correlation effects is needed.

\smallskip{}

Data availability statement: The data that supports the findings of
this study are available within the article.

\begin{acknowledgments}
\textcolor{black}{Financial support comes from MCIU/AEI/FEDER, UE
(PGC2018-097529-B-100) and }Eusko Jaurlaritza (Ref. IT1254-19)\textcolor{black}{.
The authors thank for technical and human support provided by IZO-SGI
SGIker of UPV/EHU and European funding (ERDF and ESF).}
\end{acknowledgments}


\begin{table*}
\textbf{Appendix: Theoretical and experimental equilibrium geometries}\bigskip{}
\bigskip{}

Table 1: Equilibrium bond distances ($\textrm{Å}$) of diatomic molecules
calculated at PNOF7s/cc-pVTZ level of theory. 

\qquad{}\qquad{}HF, MP2, CCSD(T) and experimental data from Ref.
\cite{nist}.
\begin{centering}
\bigskip{}
\par\end{centering}
\begin{centering}
\begin{tabular}{c|c|ccccc}
\multirow{2}{*}{Molecule} & \multirow{2}{*}{2S+1} & \multirow{2}{*}{PNOF7s} & \multirow{2}{*}{$\quad$HF$\quad$} & \multirow{2}{*}{$\quad$MP2$\quad$} & \multirow{2}{*}{CCSD(T)} & \multirow{2}{*}{$\quad$EXP$\quad$}\tabularnewline
 &  &  &  &  &  & \tabularnewline
\hline 
BeH & 2 & 1.340 & 1.339 & 1.342 & 1.350 & 1.343\tabularnewline
BO & 2 & 1.188 & 1.182 & 1.212 & 1.213 & 1.204\tabularnewline
BS & 2 & 1.600 & 1.601 & 1.613 & 1.623 & 1.609\tabularnewline
CF & 2 & 1.272 & 1.254 & 1.273 & 1.278 & 1.272\tabularnewline
CH & 2 & 1.123 & 1.106 & 1.114 & 1.123 & 1.120\tabularnewline
CN & 2 & 1.159 & 1.150 & 1.126 & 1.174 & 1.172\tabularnewline
CP & 2 & 1.550 & 1.608 & 1.518 & 1.570 & 1.562\tabularnewline
ClO & 2 & 1.595 & 1.595 & 1.593 & 1.596 & 1.570\tabularnewline
MgCl{*} & 2 & 2.190 & 2.216 & 2.203 & 2.203 & 2.199\tabularnewline
MgF{*} & 2 & 1.750 & 1.748 & 1.767 & 1.763 & 1.750\tabularnewline
MgH & 2 & 1.725 & 1.737 & 1.731 & 1.744 & 1.723\tabularnewline
NF & 3 & 1.344 & 1.292 & 1.312 & 1.320 & 1.317\tabularnewline
NH & 3 & 1.047 & 1.021 & 1.031 & 1.039 & 1.036\tabularnewline
NS & 2 & 1.509 & 1.536 & 1.423 & 1.508 & 1.494\tabularnewline
NO & 2 & 1.142 & 1.116 & 1.137 & 1.153 & 1.151\tabularnewline
OH & 2 & 0.969 & 0.951 & 0.962 & 0.971 & 0.970\tabularnewline
O$_{2}$ & 3 & 1.224 & 1.159 & 1.224 & 1.212 & 1.208\tabularnewline
PH & 3 & 1.426 & 1.414 & 1.420 & 1.429 & 1.422\tabularnewline
PO & 2 & 1.473 & 1.447 & 1.497 & 1.495 & 1.476\tabularnewline
S$_{2}$ & 3 & 1.910 & 1.872 & 1.911 & 1.916 & 1.889\tabularnewline
SH & 2 & 1.344 & 1.334 & 1.340 & 1.346 & 1.340\tabularnewline
SO & 3 & 1.442 & 1.452 & 1.501 & 1.501 & 1.481\tabularnewline
Si$_{2}$ & 3 & 2.223 & 2.223 & 2.259 & 2.180 & 2.246\tabularnewline
SiF & 2 & 1.607 & 1.596 & 1.622 & 1.620 & 1.601\tabularnewline
\end{tabular}\bigskip{}
\par\end{centering}
\centering{}{*} HF, MP2, and CCSD(T) geometries obtained using PSI4
\cite{PSI4}.
\end{table*}

\begin{table*}
Table 2: Equilibrium bond distances ($\textrm{Å}$) of polyatomic
molecules calculated at PNOF7s/cc-pVTZ level of theory. 

\qquad{}\qquad{}HF, MP2, CCSD(T) and experimental data from Ref.
\cite{nist}.

\bigskip{}

\begin{centering}
\begin{tabular}{c|c|c|ccccc}
Molecule & 2S+1 & $\quad$Bond$\quad$ & PNOF7s & $\quad$HF$\quad$ & $\quad$MP2$\quad$ & CCSD(T) & $\quad$EXP$\quad$\tabularnewline
\hline 
CH$_{2}$ & 3 & rCH & 1.091 & 1.070 & 1.074 & 1.078 & 1.085\tabularnewline
 &  & rHH & 1.997 & 1.954 & 1.968 & 1.982 & 2.008\tabularnewline
CH$_{3}$ & 2 & rCH & 1.082 & 1.071 & 1.074 & 1.079 & 1.079\tabularnewline
 &  & rHH & 1.874 & 1.855 & 1.861 & 1.869 & 1.869\tabularnewline
HCO & 2 & rCH & 1.107 & 1.108 & 1.117 & 1.121 & 1.080\tabularnewline
 &  & rCO & 1.151 & 1.151 & 1.182 & 1.182 & 1.198\tabularnewline
 &  & rHO & 2.019 & 2.019 & 2.031 & 2.038 & 1.969\tabularnewline
NH$_{2}${*} & 2 & rNH & 1.023 & 1.009 & 1.019 & 1.024 & 1.024\tabularnewline
 &  & rHH & 1.606 & 1.596 & 1.594 & 1.598 & 1.607\tabularnewline
NO$_{2}$ & 2 & rNO & 1.154 & 1.155 & 1.201 & 1.199 & 1.193\tabularnewline
 &  & rOO & 2.144 & 2.145 & 2.212 & 2.209 & 2.198\tabularnewline
OOH & 2 & rOO & 1.351 & 1.309 & 1.315 & 1.336 & 1.333\tabularnewline
 &  & rOH & 0.972 & 0.945 & 0.972 & 0.972 & 0.971\tabularnewline
 &  & rOH & 1.842 & 1.804 & 1.804 & 1.831 & 1.832\tabularnewline
PH$_{2}$ & 2 & rPH & 1.420 & 1.412 & 1.415 & 1.423 & 1.428\tabularnewline
 &  & rHH & 2.066 & 2.060 & 2.036 & 2.045 & 2.046\tabularnewline
SiH$_{3}$ & 2 & rSiH & 1.476 & 1.479 & 1.478 & 1.484 & 1.468\tabularnewline
 &  & rHH & 2.434 & 2.434 & 2.437 & 2.449 & 2.412\tabularnewline
\end{tabular}\bigskip{}
\par\end{centering}
\centering{}{*} HF, MP2, and CCSD(T) geometries obtained by using
PSI4 \cite{PSI4}.
\end{table*}

\begin{table*}
Table 3: Equilibrium bond angles (degrees) of polyatomic molecules
calculated at PNOF7s/cc-pVTZ level of theory.

\qquad{}\qquad{}HF, MP2, CCSD(T) and experimental data from Ref.
\cite{nist}.

\bigskip{}

\begin{centering}
\begin{tabular}{c|c|c|ccccc}
Molecule & 2S+1 & $\quad$Angle$\quad$ & PNOF7s & $\quad$HF$\quad$ & $\quad$MP2$\quad$ & CCSD(T) & $\quad$EXP$\quad$\tabularnewline
\hline 
CH$_{2}$ & 3 & HCH & 132.4 & 131.8 & 132.8 & 133.5 & 135.5\tabularnewline
CH$_{3}$ & 2 & HCH & 120.0 & 120.0 & 120.0 & 120.0 & 120.0\tabularnewline
HCO & 2 & HCO & 117.0 & 126.7 & 124.0 & 124.4 & 119.5\tabularnewline
NH$_{2}${*} & 2 & HNH & 103.4 & 104.5 & 102.9 & 102.6 & 103.4\tabularnewline
NO$_{2}$ & 2 & ONO & 136.6 & 136.5 & 134.1 & 134.2 & 134.1\tabularnewline
OOH & 2 & OOH & 103.7 & 105.2 & 103.1 & 103.9 & 104.3\tabularnewline
PH$_{2}$ & 2 & HPH & 93.4 & 93.7 & 92.0 & 91.9 & 91.5\tabularnewline
SiH$_{3}$ & 2 & HSiH & 111.1 & 110.7 & 111.1 & 111.2 & 110.5\tabularnewline
\end{tabular}
\par\end{centering}
\begin{centering}
\bigskip{}
\par\end{centering}
\centering{}{*} HF, MP2, and CCSD(T) geometries obtained using PSI4
\cite{PSI4}.
\end{table*}

\end{document}